\documentclass[journal]{IEEEtran}

\ifCLASSINFOpdf
\else
   \usepackage[dvips]{graphicx}
\fi
\usepackage{url}

\usepackage{graphicx}
\usepackage{amsmath,amssymb,amsfonts,amsthm}
\usepackage{hyperref}
\usepackage{tabularx}
\usepackage{verbatim}
\usepackage{caption}
\usepackage{bbm}
\usepackage{subcaption}
 \usepackage{booktabs}
\usepackage{cite}
\usepackage[switch]{lineno}
\usepackage[named]{algo}
\usepackage[noend]{algpseudocode}
\usepackage{algorithm}

\usepackage{amssymb}
\usepackage{float}
\usepackage{xcolor}

\newtheorem{theorem}{Theorem}

\newtheorem{lemma}[theorem]{Lemma}

\newtheorem{corollary}[theorem]{Corollary}

\newtheorem{remark}{Remark}

\usepackage[compact]{titlesec}
    \titlespacing{\section}{0pt}{1.0ex}{0.5ex}
    \titlespacing{\subsection}{0pt}{0.5ex}{0.5ex}
    \titlespacing{\subsubsection}{0pt}{1.0ex}{0.5ex}

\begin{document}

\title{Rao-Blackwellized Coverage Estimation in Poisson Networks: A High-Fidelity Hybrid Framework}

\author{Sunder Ram Krishnan, Junaid Farooq~\IEEEmembership{Senior Member,~IEEE}, Kumar Vijay Mishra~\IEEEmembership{Senior Member,~IEEE}, \\ 
Xingchen Liu, S. Unnikrishna Pillai~\IEEEmembership{Life Fellow,~IEEE}, and Theodore S. Rappaport~\IEEEmembership{Life Fellow,~IEEE}\vspace{-0.1in} 
\thanks{S.~R.~K. is with the Amrita Vishwa Vidyapeetham, Amritapuri, KL 690525 India. E-mail: eeksunderram@gmail.com.}
\thanks{J. F. is with the University of Michigan-Dearborn, MI 48128 USA. E-mail: mjfarooq@umich.edu.}
\thanks{K. V. M. is with the United States DEVCOM Army Research Laboratory, Adelphi, MD 20783 USA. E-mail: kvm@ieee.org.}
\thanks{X. L., S. U. P., and T. S. R. are with the New York University, Brooklyn, NY 11201 USA. E-mail: \{xl5933, pillai, ted.rappaport\}@nyu.edu. \vspace{-0.2in}}
}

\maketitle

\begin{abstract}
While stochastic geometry provides a powerful framework for the analysis of cellular networks, standard Monte Carlo simulations often suffer from slow convergence due to the stochasticity of the infinite far-field. This work introduces the \textit{Rao-Blackwellized Hybrid Estimator} (RBHE), which enhances simulation efficiency by analytically marginalizing the residual far-field interference via the conditional Laplace functional. By partitioning the interference field into $K$ dominant interferers and an infinite tail, we derive an estimator that combines exact spatial sampling with a rigorous analytical representation. We prove that the RBHE is an unbiased estimator for any finite truncation, while its systematic bias relative to the infinite-plane benchmark decays at a rate of $\mathcal{O}(K^{1-\eta/2})$. Numerical results demonstrate significant sample parsimony; in the high-reliability regime ($T = -10$ dB) with $K=2$, the RBHE yields a variance reduction gain of $90.75\times$, enabling a $98.90\%$ reduction in the spatial realizations required to reach a target precision. This framework effectively bridges the gap between tractable analytical models and high-fidelity simulations.
\end{abstract}


\IEEEpeerreviewmaketitle

\section{Introduction}
Signal coverage analysis in cellular networks is a foundational problem \cite{Andersen} typically addressed via stochastic geometry (SG) by modeling base station (BS) locations as a Poisson point process (PPP) \cite{haenggi_book_SG, ElSawy}. While SG formulations facilitate the analysis of multi-tier \cite{Dhillon} and mmWave systems \cite{heath2015millimeter}, resulting signal-to-interference-plus-noise ratio (SINR) coverage expressions often involve complex multi-dimensional integrals or frequency-domain inversions that limit interpretability and real-time optimization \cite{hamdi,Dhillon}. Efforts to modularize this include approximating interference via finite-order moments \cite{haenggi_mean_sinr, farooqprobabilistic} or the SINR meta-distribution (MD) \cite{meta_distribution, Qin}. A notable approach by Qin et al. \cite{Qin} utilizes a dominant-interferer partition to approximate the MD; however, such methods often rely on mean-field closure---approximating tail interference by its conditional mean---introducing a systematic Jensen's inequality bias and discarding far-field variance. Furthermore, exact coverage evaluations often require Gil-Pelaez inversions \cite{haenggi_book_SG} that can be numerically ill-conditioned for high SINR thresholds or non-standard path-loss models.

In this letter, we propose a modular, numerically stable \textit{Rao-Blackwellized Hybrid Estimator} (RBHE) for SINR coverage. Unlike the mean-field approximation in \cite{Qin}, our framework analytically marginalizes far-field interference via the conditional Laplace functional of the PPP, ensuring a strictly unbiased estimator with significant variance reduction over unconditioned Monte Carlo (MC) methods. Although there are semi-analytical approaches \cite{haenggi_mean_sinr}, RBHE is distinct in its use of the Rao-Blackwell theorem to provide a formal variance reduction and consequent sample parsimony guarantee. By employing real-valued Laplace transforms, the RBHE bypasses the oscillatory instabilities of complex-plane inversions. This framework is extensible to composite fading, such as $\lambda-\kappa-\mu$ \cite{wang2024lambda} or double $\lambda-\kappa-\mu$ \cite{liu2025double} distributions, where exact solutions are often intractable. %Table~\ref{tab:comparison} contrasts our contributions with existing methods \cite{Qin}. 
Specifically, our contributions are that we: 1) develop a path-loss-agnostic RBHE treating $K$ dominant interferers exactly while modeling the residual tail via its analytical Laplace functional; 2) provide a rigorous variance analysis proving a strictly positive variance reduction gain $\mathcal{G}=\mathcal{O}(K^{1-\eta})$ over raw MC; and 3) derive explicit convergence bounds showing the tail weight decays as $\mathcal{O}(K^{1-\eta/2})$, providing a criterion for choosing the truncation cardinality $K$.\\Thus, whereas \cite{Qin} primarily aims at MD approximation by treating the tail via mean-field closure thereby introducing a systematic statistical bias with no clear convergence guarantees, our goal is to achieve variance reduction using an unbiased estimator for the coverage probability modeling the tail using a Laplace functional with clear convergence bounds.
\begin{comment}
\begin{table}[ht]
\centering
\caption{Comparison of Dominant-Interferer Frameworks}
\label{tab:comparison}\footnotesize
\begin{tabular}{|l|c|c|}\hline\textbf{Feature} & \textbf{Mean-Field (Qin \cite{Qin})} & \textbf{Proposed (RBHE)} \\ \hline Statistical Bias & Systematic (Jensen's) & Unbiased (Exact) \\ \hline Tail Treatment & Mean-Field Closure & Laplace Functional \\ \hline Primary Goal & MD Approximation & Variance Reduction \\ \hline Convergence Bound & Not Provided & $\mathcal{O}(K^{1-\eta/2})$ \\ \hline Numerical Stability & Oscillatory Inversion & Real-valued Estimator \\ \hline\end{tabular}\end{table}
\end{comment}

\section{System Model and Problem Formulation}
We consider a typical user at the origin of a downlink cellular network where BS locations form a homogeneous PPP $\Phi \subset \mathbb{R}^2$ with intensity $\lambda > 0$. According to Slivnyak's Theorem \cite{Penrose}, the conditioning on a user at the origin does not alter the statistics of the remaining process, allowing us to characterize the network via the Palm distribution of the PPP. The user is served by the nearest BS $x_1 \in \Phi$, located at distance $R_1 = \|x_1\|$. Signal propagation follows a power-law path loss with exponent $\eta > 2$ and i.i.d. unit-mean exponential Rayleigh fading $\{g_i\}$. The aggregate interference is $I = \sum_{x_i \in \Phi \setminus \{x_1\}} g_i \|x_i\|^{-\eta}$. To achieve a modular characterization, we partition the interference into a \textit{dominant set} $\Phi_{\text{dom}} = \{x_2, \dots, x_K\}$ of the $K-1$ nearest interferers and a \textit{residual tail} $\Phi_{\text{tail}} = \{x_{K+1}, \dots\}$ representing the infinite far-field. The SINR is thus $\mathrm{SINR} = g_1 R_1^{-\eta} (I_{\text{dom}} + I_{\text{tail}} + \sigma^2)^{-1}$, where $I_{\text{dom}} = \sum_{i=2}^{K} g_i R_i^{-\eta}$ and $I_{\text{tail}} = \sum_{x \in \Phi_{\text{tail}}} g_x \|x\|^{-\eta}$. Our objective is to estimate the coverage probability $P_c(T): = \mathbb{P}(\mathrm{SINR} > T)$ using a Rao-Blackwellized estimator that preserves the infinite-plane nature of interference while achieving superior variance reduction over raw MC methods.

\section{The Rao-Blackwellized Hybrid Estimator}
The calculation of $P_c(T)$ involves high-dimensional spatial averaging over the PPP.\\
\textbf{Conditional Laplace Representation:} We propose a Rao-Blackwellized estimator that analytically marginalizes the fading of all interferers and the spatial randomness of the far-field tail, given the dominant geometry $\Phi_K = \{R_1, \dots, R_K\}$. By the Strong Markov Property, points in $\Phi_{\text{tail}}$ form a stationary PPP with intensity $\lambda$ restricted to the exterior of $B(0, R_K)$. For $s = T R_1^\eta$ and $g_1 \sim \text{Exp}(1)$, the conditional coverage probability is $\mathbb{P}(\mathrm{SINR} > T \mid \Phi_K) = e^{-s\sigma^2} \mathcal{L}_{\text{dom}}(s) \mathcal{L}_{\text{tail}}(s \mid R_K)$, where $\mathcal{L}_{\text{dom}}(s) = \prod_{i=2}^K (1 + s R_i^{-\eta})^{-1}$. Applying the Probability Generating Functional (PGFL), the tail Laplace transform is:
\begin{equation}
\mathcal{L}_{\text{tail}}(s \mid R_K) = \exp\left( -2\pi \lambda \int_{R_K}^{\infty} \frac{s x^{1-\eta}}{1 + s x^{-\eta}} dx \right).
\label{eq:cl}
\end{equation}
The infinite-plane coverage is then $P_{\mathrm{SG}}(T) = \mathbb{E}_{\Phi_K} [ e^{-s\sigma^2} \mathcal{L}_{\text{dom}}(s) \mathcal{L}_{\text{tail}}(s \mid R_K) ]$, which is computed via the MC estimate $\widehat{P}_c = \frac{1}{M} \sum_{m=1}^M X_m$, where $X_m$ is the term inside the expectation for the $m$-th trial. We term this the RBHE as it marginalizes over fast-fading and spatial tail uncertainty. While $\Phi_K$ is not a sufficient statistic for the infinite realization $\Phi$, the estimator is unbiased and converges to the full Rao-Blackwellized estimator as $K \to \infty$.

\subsection{Analytical Solutions and Path-Loss Agnosticism}
The integral in \eqref{eq:cl} is valid for any $\eta > 2$ and can be solved in terms of the Gauss hypergeometric function ${}_2F_1$. This provides a unified framework that does not require symbolic re-derivation for specific network parameters:\small
\begin{equation}
\mathcal{L}_{\text{tail}}(s \mid R_K) = \exp\left( -\pi \lambda R_K^2 \left[ {}_2F_1\left(1, \frac{2}{\eta}; 1+\frac{2}{\eta}; -s R_K^{-\eta}\right) - 1 \right] \right).
\label{eq:general_L}
\end{equation}\normalsize
For the case of $\eta=4$, we get an elegant arctangent form:
\begin{equation}
\mathcal{L}_{\text{tail}}(s \mid R_K)\Big|_{\eta=4} = \exp\left( -\pi\lambda \sqrt{s} \left[ \frac{\pi}{2} - \arctan\left(\frac{R_K^2}{\sqrt{s}}\right) \right] \right).
\label{eq:eta4_L}
\end{equation}

\subsection{Series Representation and Asymptotic Behavior}
To gain insight into the convergence of the tail approximation, we provide a series expansion of the log-Laplace functional. 

\begin{lemma}[Multiscale Series Representation]
For $s R_K^{-\eta} < 1$, the log-Laplace functional of the interference tail, $\Psi_{\text{tail}}(s) = \ln \mathcal{L}_{\text{tail}}(s \mid R_K)$, admits a cumulant-based series expansion:\footnotesize
\begin{align}
\Psi_{\text{tail}}(s) = \underbrace{-s \left( \frac{2\pi\lambda R_K^{2-\eta}}{\eta - 2} \right)}_{\text{Mean-Field Term}} &+ \underbrace{\frac{s^2}{2} \left( \frac{2\pi\lambda R_K^{2-2\eta}}{\eta - 1} \right)}_{\text{Variance-Correction}}\nonumber\\ &+ \sum_{m=3}^{\infty} \frac{2\pi\lambda (-s)^m R_K^{2-m\eta}}{m\eta - 2}.
\label{eq:series}
\end{align}
\label{lem:ser}
\end{lemma}\normalsize

\begin{proof}
$\Psi_{\text{tail}}(s)$ is the conditional cumulant generating function of $I_{\text{tail}}$, admitting the Taylor expansion $\sum_{m=1}^{\infty} \frac{(-s)^m}{m!} \kappa_m(R_K)$. Term-wise integration of the series expansion of the exponent in the PGFL (cf. \eqref{eq:cl}) yields the general term $\frac{2\pi\lambda (-s)^m R_K^{2-m\eta}}{m\eta - 2}$. 1) For $m=1$, the first term is $-s \kappa_1(R_K)$; identifying $\kappa_1(R_K) = \frac{2\pi\lambda R_K^{2-\eta}}{\eta-2}$ recovers the conditional mean $\mathbb{E}[I_{\text{tail}} \mid R_K]$ via Campbell's Theorem. 2) For $m=2$, the second term $\frac{s^2}{2!} \kappa_2(R_K)$ yields $\kappa_2(R_K) = \frac{2\pi\lambda R_K^{2-2\eta}}{\eta-1}$, which is the conditional variance $\mathrm{Var}(I_{\text{tail}} \mid R_K)$ capturing far-field spatial fluctuations.
\end{proof}

This series expansion demonstrates that the dominant contribution of the tail scales as $\mathcal{O}(R_K^{2-\eta})$, providing a clear physical intuition for the interference decay.

\begin{remark}[Mathematical Distinction from Mean-Field Approximations]
It is instructive to contrast the proposed RBHE with existing dominant-interferer approximations, such as the Lambert-W based approach in \cite{Qin}. In \cite{Qin}, the tail interference is approximated by its conditional mean by ignoring the higher-order moments of the far-field. By utilizing the exact Laplace functional $\mathcal{L}_{\text{tail}}(s \mid R_K)$, the RBHE preserves the full expansion of the interference distribution; see \eqref{eq:series}.
%While \cite{Qin} relies on the first cumulant (mean), the RBHE utilizes the complete CGF, thereby eliminating the approximation bias inherent in mean-field methods.
While \cite{Qin} uses truncation at the Mean-Field Term ($m=1$) of \eqref{eq:series} to enable a transcendental solution for the MD, the proposed RBHE achieves its superior stability by analytically integrating the Variance-Correction ($m=2$) and all higher-order cumulants ($m \geq 3$) of the infinite-plane interference.
\end{remark}

\begin{algorithm}
\caption{Rao-Blackwellized Hybrid Estimator (RBHE)}
\begin{algorithmic}[1]
\State \textbf{Input:} Density $\lambda$, exponent $\eta$, noise $\sigma^2$, threshold $T$, truncation $K$, trials $M$.
\State \textbf{Output:} Coverage Estimate $\widehat{P}_c$.
\For{$m = 1$ to $M$}
    \State Sample $K$ independent variables $E_1, \dots, E_K \sim \text{Exp}(\pi \lambda)$.
    \State Compute ordered squared distances: $R_i^2 = \frac{1}{\pi \lambda} \sum_{j=1}^i E_j$ for $i=1, \dots, K$.
    \State Set transform parameter: $s = T (R_1^2)^{\eta/2}$.
    \State Evaluate dominant product: $\mathcal{L}_{\text{dom}} = \prod_{i=2}^K (1 + s (R_i^2)^{-\eta/2})^{-1}$.
    \State Compute tail functional $\mathcal{L}_{\text{tail}}(s \mid R_K)$ using \eqref{eq:general_L} or \eqref{eq:eta4_L}.
    \State Store sample: $X_m = e^{-s\sigma^2} \cdot \mathcal{L}_{\text{dom}} \cdot \mathcal{L}_{\text{tail}}$.
\EndFor
\State \textbf{Return:} $\widehat{P}_c = \frac{1}{M} \sum_{m=1}^M X_m$.
\end{algorithmic}
\end{algorithm}

\section{Residual Tail Weight and Variance Reduction}
 Define \(\delta_K := \mathbb{E}_{\Phi_K} \left[ 1 - \mathcal{L}_{\text{tail}}(s \mid R_K) \right]\).
The characterization of the residual tail weight $\delta_K$ serves a dual purpose in validating the RB-Hybrid framework. 
First, it provides a theoretically grounded selection criterion for the truncation cardinality $K$. By proving below that $\delta_K$ decays at a rate of $\mathcal{O}(K^{1-\eta/2})$, we demonstrate that the dependence of the estimator on the analytical tail approximation---rather than on the exact stochastic geometry of the dominant set---diminishes rapidly as $K$ increases. 
Second, $\delta_K$ quantifies the representational gap between finite-network simulations and infinite-plane models. In standard truncated simulations, $\delta_K$ represents the systematic bias (truncation error) incurred by ignoring the far-field interference. In our framework, however, $\delta_K$ represents the fraction of the aggregate interference field that is stabilized via Rao-Blackwellization. 
By bounding $\delta_K$, we ensure that the hybrid approach is not merely a numerical heuristic, but a convergent estimator that recovers the infinite-plane characteristic with a controllable and vanishing error floor.

\begin{theorem}[Total Estimator Convergence]
\begin{enumerate}
 \item The total deviation satisfies $| \widehat{P}_c - P_{\mathrm{SG}}(T) | \leq \epsilon_M$, where the statistical error $\epsilon_M$ has the asymptotic property $\sqrt{M}\epsilon_M \xrightarrow{d} \mathcal{N}(0, V_K)$ as $M \to \infty$, and $V_K = \mathrm{Var}(X_m)$ is the finite variance of the RB-Hybrid samples. 
 \item For all large \(K>K_0(\eta)\), we have:
\begin{equation}
\delta_K \leq \frac{4 T \Gamma(\frac{\eta}{2}+1)}{\eta-2} K^{1-\eta/2},
\label{eq:dc}
\end{equation}
which implies $\delta_K \to 0$ as $K \to \infty$ for any $\eta > 2$.
\label{thm:total_convergence}
\end{enumerate}
\end{theorem}

\begin{proof}
1) %The statistical error $\epsilon_M$ represents the fluctuation of the empirical mean. 
Since $X_m$ is a product of Laplace transforms, $X_m \in [0, 1]$, and thus $V_K$ is bounded. 
%By the Weak Law of Large Numbers, $\widehat{P}_c \xrightarrow{p} P_{\mathrm{SG}}(T)$ as $M \to \infty$. 
%By the Central Limit Theorem, the distribution of $\epsilon_M$ is asymptotically normal with a standard deviation scaling as $1/\sqrt{M}$. Thus, $\epsilon_M = \mathcal{O}_p(M^{-1/2})$.
The claim follows from the Central Limit Theorem.

2) By applying the inequality $1-e^{-x} \leq x$ to the Laplace transform of the tail interference, we bound the residual as:
\begin{equation*}
\delta_K \leq \mathbb{E} \left[ 2\pi\lambda \int_{R_K}^{\infty} \frac{s x}{x^\eta + s} dx \right] < \frac{2\pi\lambda T}{\eta-2} \mathbb{E} \left[ R_1^\eta R_K^{2-\eta} \right].
\end{equation*}
The joint PDF of the $1^{\text{st}}$ and $K^{\text{th}}$ nearest neighbor distances $R_1$ and $R_K$ in a 2D PPP with density $\lambda$ is given by:
\begin{align*}
f_{R_1, R_K}(r_1, r_K) =& \frac{4(\pi\lambda)^K}{(K-2)!} r_1 r_K (r_K^2 - r_1^2)^{K-2} e^{-\pi\lambda r_K^2}, \\ & 0 \leq r_1 \leq r_K < \infty.
\end{align*}
We evaluate the expectation $\mathcal{I} = \mathbb{E}[R_1^\eta R_K^{2-\eta}]$ via the double integral:\footnotesize
\begin{equation*}
\mathcal{I} = \int_0^\infty \int_0^{r_K} r_1^\eta r_K^{2-\eta} \frac{4(\pi\lambda)^K}{(K-2)!} r_1 r_K (r_K^2 - r_1^2)^{K-2} e^{-\pi\lambda r_K^2} dr_1 dr_K,
\end{equation*}\normalsize
substituting $u = (r_1/r_K)^2$.
The inner integral over $r_1$ yields:
\[
\int_0^{r_K} r_1^{\eta+1} (r_K^2 - r_1^2)^{K-2} dr_1 =\frac{1}{2} r_K^{\eta + 2K - 2} \frac{\Gamma(\frac{\eta}{2}+1)\Gamma(K-1)}{\Gamma(\frac{\eta}{2}+K)}.
\]
Substituting this back into the outer integral over $r_K$, we get \(\mathcal{I}=\frac{\Gamma(\frac{\eta}{2}+1)}{\pi\lambda} \frac{\Gamma(K+1)}{\Gamma(K+\frac{\eta}{2})}\).
\begin{comment}
\begin{align*}
\int_0^{r_K} r_1^{\eta+1} (r_K^2 - r_1^2)^{K-2} dr_1 &= \frac{1}{2} r_K^{\eta + 2K - 2} \int_0^1 u^{\eta/2} (1-u)^{K-2} du  \\
&= \frac{1}{2} r_K^{\eta + 2K - 2} \frac{\Gamma(\frac{\eta}{2}+1)\Gamma(K-1)}{\Gamma(\frac{\eta}{2}+K)}.
\end{align*}
Substituting this back into the outer integral over $r_K$:
\begin{align*}
\mathcal{I} &= \frac{2(\pi\lambda)^K}{(K-2)!} \frac{\Gamma(\frac{\eta}{2}+1)\Gamma(K-1)}{\Gamma(\frac{\eta}{2}+K)} \int_0^\infty r_K^{2K+1} e^{-\pi\lambda r_K^2} dr_K  \\
&= \frac{2(\pi\lambda)^K}{(K-2)!} \frac{\Gamma(\frac{\eta}{2}+1)\Gamma(K-1)}{\Gamma(\frac{\eta}{2}+K)} \left( \frac{K!}{2(\pi\lambda)^{K+1}} \right)  \\
&= \frac{\Gamma(\frac{\eta}{2}+1)}{\pi\lambda} \frac{\Gamma(K+1)}{\Gamma(K+\frac{\eta}{2})}.
\end{align*}
\end{comment}
Using (Wendal's) limit \cite{abra} (p. 257, 6.1.46), $\frac{\Gamma(K+1)}{\Gamma(K+\frac{\eta}{2})} \leq 2K^{1-\eta/2}$ for all large \(K\), and substituting $\mathcal{I}$ back into the bound for $\delta_K$:
\begin{equation*}
\delta_K \leq \frac{4\pi\lambda T}{\eta-2} \cdot \frac{\Gamma(\frac{\eta}{2}+1)}{\pi\lambda} K^{1-\eta/2} = \frac{4 T \Gamma(\frac{\eta}{2}+1)}{\eta-2} K^{1-\eta/2},
\end{equation*}
completing the proof.
\end{proof}

As alluded to, our idea behind the RBHE is to achieve variance reduction over full MC. The following result seeks to quantify the reduction in variance achieved using the tail functional $\mathcal{L}_{\text{tail}}(s \mid R_K)$.

\begin{theorem}[Variance Reduction Gain]
Let $\hat{P}_{MC} = e^{-s(\sigma^2+I_{\text{dom}} + I_{\text{tail}})}$ be the crude MC estimator viewed as a random variable over the joint space of BS locations and fading. Furthermore, let $\widetilde{P}_c = \mathbb{E}[\hat{P}_{MC} \mid \Phi_K]$ be the corresponding Rao-Blackwellized estimator. The variance reduction gain $\mathcal{G}$ is given by:
\begin{align}
&\mathrm{Var}(\hat{P}_{MC}) = \mathrm{Var}(\widetilde{P}_c) \nonumber\\&+ \underbrace{\mathbb{E}_{\Phi_K} \left[ e^{-2s\sigma^2} \mathcal{L}_{\text{dom}}(2s) \left( \mathcal{L}_{\text{tail}}(2s \mid R_K) - \mathcal{L}_{\text{tail}}(s \mid R_K)^2 \right) \right]}_{\mathcal{G} \geq 0}.\label{eq:vg}
\end{align}
\label{thm:vg}
\end{theorem}

\begin{proof}
By the Law of Total Variance:
\begin{equation*}
\mathrm{Var}(\hat{P}_{MC}) = \mathrm{Var}(\mathbb{E}[\hat{P}_{MC} \mid \Phi_K]) + \mathbb{E}[\mathrm{Var}(\hat{P}_{MC} \mid \Phi_K)].
\end{equation*}
The first term is exactly $\mathrm{Var}(\widetilde{P}_c)$. The second term, representing the variance reduction $\mathcal{G}$, is:\small
\begin{align*}
\mathcal{G} &= \mathbb{E}_{\Phi_K} \left[ \mathbb{E}_{\Phi_{\text{tail}}} [ \hat{P}_{MC}^2 \mid \Phi_K ] - (\mathbb{E}_{\Phi_{\text{tail}}} [ \hat{P}_{MC} \mid \Phi_K ])^2 \right]  \\
&= \mathbb{E}_{\Phi_K} \left[ e^{-2s\sigma^2} \mathcal{L}_{\text{dom}}(2s) \left( \mathcal{L}_{\text{tail}}(2s \mid R_K) - \mathcal{L}_{\text{tail}}(s \mid R_K)^2 \right) \right].
\end{align*}\normalsize
\begin{comment}
Recognizing the terms as conditional Laplace functionals:
\begin{equation*}
\mathcal{G} = \mathbb{E}_{\Phi_K} \left[ e^{-2s\sigma^2} \mathcal{L}_{\text{dom}}(2s) \left( \mathcal{L}_{\text{tail}}(2s \mid R_K) - \mathcal{L}_{\text{tail}}(s \mid R_K)^2 \right) \right].
\end{equation*}
\end{comment}
By Jensen's Inequality, the term in the parenthesis is non-negative for every $R_K$. Thus, $\mathrm{Var}(\widetilde{P}_c) \leq \mathrm{Var}(\hat{P}_{MC})$, where the reduction $\mathcal{G}$ is the variance of the tail interference that is smoothed out by the analytical marginalization.
\end{proof}

\begin{corollary}[Asymptotic Variance Gain]
For large $K$ such that $R_K > (2s)^{1/\eta}$, the variance reduction gain $\mathcal{G}$ is bounded by:
\begin{equation}
\mathcal{G} \leq \frac{2\pi\lambda T^2 \mathbb{E}[R_1^{2\eta} R_K^{2-2\eta}]}{\eta - 1}= \mathcal{O}(K^{1-\eta}).
\end{equation} 
\end{corollary}

In the context of a PPP, for any finite threshold $T$, the condition $R_K > (2s)^{1/\eta}$ holds with probability approaching 1 as $K \to \infty$.

\begin{proof}
The variance reduction gain is defined by the expectation of the conditional variance $\Delta \mathcal{L} = \mathcal{L}_{\text{tail}}(2s \mid R_K) - \mathcal{L}_{\text{tail}}(s \mid R_K)^2$. To bound this exactly, we express the Laplace functional in its exponentiated PGFL form:
\begin{equation*}
\Delta \mathcal{L} = e^{\Psi_{\text{tail}}(2s)} - e^{2\Psi_{\text{tail}}(s)} = e^{2\Psi_{\text{tail}}(s)} \left( e^{\Psi_{\text{tail}}(2s) - 2\Psi_{\text{tail}}(s)} - 1 \right).
\end{equation*}
Using the inequality $e^x - 1 \leq x e^x$ for $x \geq 0$, and noting that $e^{\Psi_{\text{tail}}(2s)} \leq 1$ for large \(K\) under consideration, we focus on the exponent difference. From the series representation in Lemma \ref{lem:ser}:
\begin{equation*}
\Psi_{\text{tail}}(2s) - 2\Psi_{\text{tail}}(s) = \sum_{m=1}^{\infty} \frac{2\pi\lambda (-s)^m R_K^{2-m\eta}}{m\eta - 2} (2^m - 2).
\end{equation*}
The $m=1$ term vanishes identically. For $\eta > 2$ and large $R_K$, the series is dominated by the $m=2$ term. Invoking the Alternating Series Remainder Theorem, which is applicable provided $R_K > (2s)^{1/\eta}$:\small
\begin{equation*}
\Delta \mathcal{L} \leq \Psi_{\text{tail}}(2s) - 2\Psi_{\text{tail}}(s) \leq \frac{2\pi\lambda s^2 R_K^{2-2\eta}}{2\eta - 2} (2^2 - 2) = \frac{4\pi\lambda s^2 R_K^{2-2\eta}}{2\eta - 2}.
\end{equation*}\normalsize
Substituting $s = T R_1^\eta$, we have \(\mathcal{G} = \mathbb{E}_{\Phi_K}[e^{-2s\sigma^2} \mathcal{L}_{\text{dom}}(2s) \Delta \mathcal{L}]\leq \mathbb{E}_{\Phi_K}[\Delta \mathcal{L}].\)
\begin{comment}
and simplifying the denominator:
\begin{equation*}
\Delta \mathcal{L} \leq \frac{2\pi\lambda T^2 R_1^{2\eta} R_K^{2-2\eta}}{\eta - 1}.
\end{equation*}
The gain is $\mathcal{G} = \mathbb{E}_{\Phi_K}[e^{-2s\sigma^2} \mathcal{L}_{\text{dom}}(2s) \Delta \mathcal{L}]$. Since the product of Laplace functionals is bounded by unity, we have $\mathcal{G} \leq \mathbb{E}_{\Phi_K}[\Delta \mathcal{L}]$. 
\end{comment}
The expectation $\mathbb{E}[R_1^{2\eta} R_K^{2-2\eta}]$ can be evaluated as in Theorem \ref{thm:total_convergence},
yielding $\mathcal{G} = \mathcal{O}(K^{1-\eta})$, completing the proof.
\end{proof}
\begin{comment}
Comparing the rates of decay, we observe a fundamental scaling law:
\(
\frac{\delta_K}{\sqrt{\mathcal{G}}} \propto K^{1/2}.
\)
This result reveals that the stochastic uncertainty $\mathcal{G}$ vanishes at a superior rate of $\mathcal{O}(K^{1-\eta})$ compared to the systematic bias $\delta_K \sim \mathcal{O}(K^{1-\eta/2})$. This disparity theoretically justifies the RBHE framework: as $K$ increases, the estimator's variance is suppressed quadratically faster than the mean-field truncation error, ensuring that the tail noise is effectively marginalized even for moderate values of $K$.
\end{comment}

\begin{remark}[Systematic Selection of \(K\)]
The derived scaling laws provide a formal basis for selecting the truncation cardinality \(K\). To ensure the systematic bias \(\delta_K\) is bounded by a target precision \(\epsilon\), the condition
\begin{equation*}
K \geq \left( \frac{4 T \Gamma(\frac{\eta}{2}+1)}{\epsilon (\eta-2)} \right)^{\frac{2}{\eta-2}}
\end{equation*}
serves as a sufficient, albeit conservative, upper bound (cf. \eqref{eq:dc}). While this expression suggests that \(K\) must scale linearly with the threshold \(T\) (for \(\eta=4\)), in practice, the bias is further suppressed by the dominant interference. Specifically, the term \(e^{-sI_{\text{dom}}}\)
in the estimator effectively masks the tail truncation error, ensuring high fidelity even for \(K\) significantly smaller than the theoretical bound. For instance, at \(T=0\) dB and \(\eta=4\), setting \(K=20\) yields a conservative theoretical precision of \(\epsilon =0.2\), though numerical evaluations in the next section typically reveal much higher accuracy. Since the variance reduction gain \(\mathcal{G}= \mathcal{O}(K^{1-\eta})\) vanishes quadratically faster than the bias, the RBHE ensures that stochastic uncertainty is marginalized effectively even for moderate \(K\), drastically reducing simulation overhead compared to large-scale finite-window MC methods.
\end{remark}

These results demonstrate that the RB-Hybrid approach provides a mathematically controlled approximation of the infinite network, where the degree of precision can be explicitly tuned via the parameter $K$. 

\section{Numerical Experiments and Discussion}

In this section, we validate the proposed RBHE by assessing its convergence to the infinite-plane benchmark and quantifying its statistical efficiency relative to crude MC sampling.

\subsection{Experimental Setup}
To ensure reproducibility, we model a downlink cellular network where BS locations follow a homogeneous PPP $\Phi$ with density $\lambda = 1 \text{ BS/km}^2$. The path-loss exponent is set to $\eta = 4$, and the additive noise power is $\sigma^2 = 0.1$. The SINR threshold $T$ is varied from $-10$ to $20$ dB.

The simulation utilizes the inter-arrival property of the distance-squared process as in Algorithm 1. For each realization, the sequence of squared distances $\{R_k^2\}$ is generated via the cumulative sum of independent exponential random variables with rate $\pi \lambda$:
\(
R_k^2 = \frac{1}{\pi\lambda}\sum_{i=1}^{k} E_i, \quad E_i \sim \text{Exp}(\pi \lambda).
\)
This method allows for the exact representation of an infinite-plane network, thereby eliminating the need for a finite simulation window and avoiding associated edge-effect biases. For the crude MC baseline, we simulate the nearest $100$ interferers to approximate the infinite interference field. For the RBHE, we consider truncation cardinalities $K \in \{2, 5, 20\}$ and analytically marginalize the infinite tail interference $I_{\text{tail}}$ using the Laplace functional in \eqref{eq:general_L} (or \eqref{eq:eta4_L}). Both estimators are evaluated using $M = 10,000$ independent spatial realizations for each data point to ensure statistical stability in the high-threshold regime.

\begin{figure}[t]
\centering
\includegraphics[width=1.0\columnwidth]{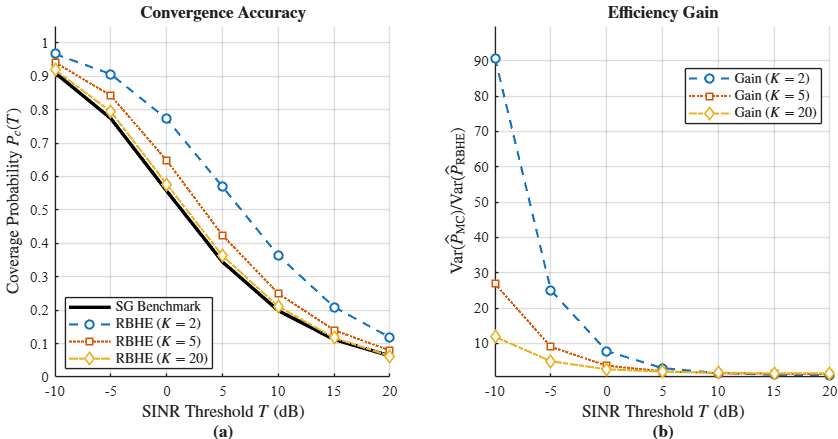}
\caption{Performance characterization of the proposed RBHE ($\eta = 4, \lambda = 1$ BS/km$^2$): (a) Convergence of the coverage probability $P_c(T)$ to the exact SG benchmark as the dominant set cardinality $K$ increases; (b) Variance reduction gain $\mathcal{G}_{\text{ratio}}$ relative to crude MC sampling, demonstrating the sample parsimony achieved by marginalizing the far-field interference tail.}
\label{fig:gain}
\end{figure}

\subsection{Convergence and Estimator Fidelity}
Fig. \ref{fig:gain}(a) illustrates the SINR coverage probability $P_{c}(T)$. The results demonstrate that the RBHE converges monotonically to the exact SG benchmark as $K$ increases. Consistent with Theorem \ref{thm:total_convergence}, the systematic bias $\delta_K$---arising from the discretization of the dominant interferers---decays at a rate of $\mathcal{O}(K^{1-\eta/2})$. Empirically, we observe that for $K=20$, the RBHE recovers the exact infinite-limit statistics with high fidelity, effectively closing the representational gap and providing a smooth, continuous characterization of the coverage surface.

\subsection{Statistical Efficiency and Sample Parsimony}
The statistical advantage of the RBHE is characterized by its \textit{sample parsimony}---the ability to achieve a target precision with significantly fewer spatial realizations $M$ than crude MC. This is quantified by the variance reduction gain $\mathcal{G}_{\text{ratio}} = \mathrm{Var}(\hat{P}_{\text{MC}}) / \mathrm{Var}(\hat{P}_{\text{RBHE}})$, depicted in Fig. \ref{fig:gain}(b). The empirical results confirm the variance decomposition analyzed in Theorem \ref{thm:vg}, where the elimination of conditional variance in the tail leads to substantial efficiency gains. 

As summarized in Table \ref{tab:savings}, the RBHE provides its most dramatic impact in the high-reliability regime. At $T = -10$ dB with $K=2$, the RBHE achieves a variance reduction gain of $90.75\times$, implying that the estimator requires approximately $98.9\%$ fewer spatial realizations than crude MC to reach the same target standard error. Even in the interference-limited regime ($T=10$ dB) with a larger dominant set ($K=20$), the RBHE maintains a gain of $1.70\times$, saving over $41\%$ of the required realizations. These results prove that the RBHE is a mathematically rigorous and computationally superior alternative for large-scale network evaluations.

\begin{table}[ht]
\centering
\caption{Statistical Efficiency and Resource Savings for $\eta=4$ ($M=10,000$)}
\label{tab:savings}
\begin{tabular}{@{}cccc@{}}
\toprule
\textbf{Threshold $T$} & \textbf{Cardinality $K$} & \textbf{Variance Gain $\mathcal{G}_{\text{ratio}}$} & \textbf{Realizations Saved} \\ \midrule
$-10$ dB & $2$  & $90.75$ & $98.90\%$ \\
$-5$ dB  & $2$  & $25.06$ & $96.01\%$ \\
$0$ dB   & $5$  & $3.71$  & $73.05\%$ \\
$5$ dB   & $5$  & $2.11$  & $52.58\%$ \\
$10$ dB  & $20$ & $1.70$  & $41.18\%$ \\
$15$ dB  & $20$ & $1.53$  & $34.64\%$ \\
$20$ dB  & $20$ & $1.56$  & $35.90\%$ \\ \bottomrule
\end{tabular}
\end{table}

\section{Conclusion}
In this letter, we developed and validated the RBHE for the efficient evaluation of coverage probability in Poisson cellular networks. By marginalizing the stochasticity of the infinite interference tail, the RBHE provides a strictly unbiased estimator with a formal variance reduction guarantee. Our theoretical analysis established that the systematic bias relative to the infinite-plane benchmark decays as $\mathcal{O}(K^{1-\eta/2})$, while the stochastic uncertainty is suppressed at a superior rate of $\mathcal{O}(K^{1-\eta})$. 

The practical utility of the RBHE was demonstrated through extensive simulations, revealing a variance reduction gain of $90.75\times$ for $K=2$ at low SINR thresholds. This gain translates to a $98.90\%$ reduction in the computational realizations required to reach a target precision, effectively transforming rare-event simulation into a computationally efficient task. Future work will extend this framework to multi-tier heterogeneous networks and non-Poissonian BS deployments, where analytical solutions for the interference tail are conventionally considered intractable.

\clearpage
\bibliographystyle{IEEEtran}
\bibliography{refs}
\end{document}